\documentclass[aps,twocolumn,amsmath,amssymb]{revtex4}

\usepackage{amssymb, amsmath}
\usepackage{graphicx}
\usepackage{dcolumn}
\usepackage{bm}

\newcommand{\unsw}{School of Mathematics and Statistics, The University of New South Wales, Sydney NSW 2052, Australia}
\newcommand{\victoria}{School of Earth and Ocean Sciences, University of Victoria, Victoria, British Columbia, Canada}
\newcommand{\degree}{\ensuremath{^\circ}}

\begin{document}

\title{Optimally coherent sets in geophysical flows: A new approach to delimiting the stratospheric polar vortex}

\author{Naratip Santitissadeekorn$^1$,
Gary Froyland$^1$,
and Adam Monahan$^2$\\ \quad
}

\affiliation{$^1$ \unsw \\
$^2$ \victoria}


\date{\today}

\begin{abstract}
The ``edge'' of the Antarctic polar vortex is known to behave as a
barrier to the meridional (poleward) transport of ozone during the
austral winter. This chemical isolation of the polar vortex from the
middle and low latitudes produces an ozone minimum in the vortex
region, intensifying the ozone hole relative to that which would be
produced by photochemical processes alone.
Observational determination of the vortex edge remains an active field of research.  In this letter, we obtain objective estimates of the structure of the polar
vortex by introducing a new technique based on
transfer operators that aims to find regions with minimal external
transport. Applying this new technique to European Centre for
Medium-Range Weather Forecasts (ECMWF) ERA-40 three-dimensional
velocity data we produce an improved three-dimensional estimate of
the vortex location in the upper stratosphere where the vortex is most pronounced. This novel computational approach has wide
potential application in detecting and analysing mixing structures
in a variety of atmospheric, oceanographic, and general fluid
dynamical settings.
\end{abstract}
\maketitle

\section{\label{sec:Introduction}Introduction}
Stratospheric ozone in the Southern Hemisphere high latitudes has
decreased dramatically since the early 1970s.  This long-term trend
has been attributed to a combination of natural and anthropogenic
factors~\citep{McIntyre89,Webster93,Shepherd07, Shepherd08}. In
particular, it has been discovered that the ozone depletion in the
lower stratosphere of the Southern Hemisphere is particularly
pronounced, due in part to a strong barrier to meridional transport
between middle and high latitudes during the austral winter and
early spring~\citep{McIntyre89}. Barriers such as these, which often coexist with turbulent mixing, play a major role in the dynamics of the stratosphere. The polar vortex is a known strong barrier to transport, enclosing a persistent, non-dispersive, coherent region over the high latitudes.
Our aim in this letter is to precisely determine the
spatial location and movements of this coherent region, improving
significantly over existing methods of estimation.
Our study focuses on the upper stratosphere where the polar vortex is best developed.




It is common meteorological practice to diagnose the polar vortex
edge at the position of maximum meridional gradient of potential
vorticity (PV). Potential vorticity is a quantity combining measures
of circulation and stratification which is materially conserved for
adiabatic, inviscid flow (both of which are good approximations in
stratospheric flow over timescales of a week or two). It can be
shown that strong PV gradients produce a ``restoring force''
inhibiting meridional motion of air parcels~\cite{Vallis06}.
Nevertheless, PV gradients alone provide only indirect measures of
mixing barriers. In contrast, the present study characterises
regions of minimal mixing directly in terms of the transport
properties of the observed stratospheric flow. We present an
innovative new mathematical technique to determine the polar vortex
location at different times, {directly} as coherent structures in
observed velocity fields. Lagrangian PV-based measures of the
vortex such as those presented in \cite{Sobel97} are complicated by the fact that PV is generally a noisy
field (as vorticity is the curl of the velocity field). The velocity
field is generally much smoother; barriers to mixing estimated from
the velocity field can be expected to be less sensitive to
(poorly-observed) small-scale features of the flow.




Our new mathematical approach for detecting minimal transport
structures with high accuracy has a broad range of potential
applications to geophysical fluids. For example, transport
properties in other long-lived atmospheric coherent structures such
as blocking highs are of interest. There is also increasing interest
in the transport properties of mesoscale (on the order of 10 to 100
km in diameter) ocean eddies and their influence on biological
processes within the upper, sunlit part of the water
column~\cite{CBT07,NKDR08}.

\section{Input data and non-autonomous flow}
Our input data consists of three-dimensional velocity fields obtained from the ECMWF
ERA-40 data set (http://data.ecmwf.int/data/index.html).
The data is on a three-dimensional grid with 2.5 degree resolution in the latitude and longitude direction (144 by 73 grid points over the Southern Hemisphere).  Vertical coordinates are in units of hPa, with data provided at 5 pressure levels (3, 5, 7, 10, 20 hPa).
We use 62 days of 6-hourly velocity fields from August 1 to September 31
in 1999.
The velocity fields will be interpolated linearly in space and in time; thus we can only aim to detect features at the resolution of the data provided.
While we recognize that there may be biases in the reanalysis data, particularly in the upper pressure levels near the model's upper boundary, the purpose of
this study is to demonstrate the ability of the transfer operator approach to characterize coherent sets in highly unsteady flows. A climatology of the polar
vortex would require a more careful consideration of the dataset under consideration

Our interest is focused on the Lagrangian dynamics
in the higher latitudes of Southern Hemisphere. Therefore, we will work on the phase space
$X= S^{1}\times [-90\degree, -30\degree]\times D$, where $S^{1}$ is
a circle parameterized from $0\degree$ to $360\degree$ and
$D=[3,20]$ denotes the range of pressure in hPa. The Lagrangian
motion of passive tracers is represented by their trajectories
$x(t):=\Phi(x,t; \tau)$, where the flow map
$\Phi:X\times\mathbb{R}\times\mathbb{R}\to X$ is a function of time
$\tau$ and gives the terminal point $\Phi(x,t;\tau)$ in $X$ of a
particle initially located at $x\in X$ at time $t$ and flowing for
$\tau$ time units. The flow map $\Phi(x_0,t_0; \tau)$ is obtained as
a solution of the nonautonomous ODE $\frac{dx}{dt}=f(x(t),t)$ with
initial condition $x(t_0)=\Phi(x_0,t_0;0)$, where $f(x(t),t)$ in
this report is the prescribed velocity data.

\section{Almost-invariant sets, coherent pairs, and transfer operators}
We shall be interested in finding a pair of sets $A_t, A_{t+\tau}$
at times $t$ and $t+\tau$ so that
$\Phi(A_{t+\tau},t+\tau;-\tau)\approx A_t$.  Moreover, this pair of
sets should retain this property even when some diffusion is added
to the system. Let $\mu$ be a probability measure that is preserved
by the flow at all times. We call $A_t, A_{t+\tau}$ a
\emph{$(\rho_0,t,\tau)-$coherent pair} if
\begin{equation}\label{eq:invariant}
\rho_\mu(A_t,A_{t+\tau}):={\mu(A_t\cap \Phi(A_{t+\tau},t+\tau;
-\tau))}/{\mu(A_t)}\ge \rho_0,
\end{equation}
and $\mu(A_t)=\mu(A_{t+\tau})$.
The condition on addition of diffusion is crucial. Clearly, there
are many $(\rho_0,t,\tau)$-coherent pairs according to the above
definition without diffusion. One may simply select any set
$A_t\subset X$ and define $A_{t+\tau}=\Phi(A_t,t;\tau)$ to produce a
$(1,t,\tau)$-coherent pair. In chaotic systems, such an image set
$A_{t+\tau}$ is likely to be significantly less regular than $A_t$
because of stretching and folding.  We are seeking
$(\rho_0,t,\tau)$-coherent pairs with \emph{both} sets regular.  The
requirement that (\ref{eq:invariant}) hold even under diffusion acts
as a selection principle, removing irregular sets, and selecting
pairs that are robust to perturbation. At a certain level of
diffusion, we may ask to find a coherent pair that maximises
$\rho_0$, and may expect a unique such pair.

To identify sets satisfying~\eqref{eq:invariant},
we use a transfer operator
$\mathcal{P}_{t,\tau}:L^1(X,m)\circlearrowleft$ defined by
\begin{equation}\label{eq:pfeqn}
\mathcal{P}_{t,\tau}g(x):=\frac{g(\Phi(x,t+\tau;-\tau))}{|\det
D\Phi(\Phi(x,t+\tau;-\tau),t; \tau)|}
\end{equation}
where $m$ is the normalized Lebesgue measure on $X$. In particular,
if $g(x)$ is a density of passive tracers at time $t$,
$\mathcal{P}_{t,\tau}g(x)$ provides their density at time $t+\tau$
induced by the flow $\Phi$.

In the autonomous setting,
eigenfunctions $f$ of $\mathcal{P}_{t,\tau}$ ($=\mathcal{P}_\tau$
for all $t$) corresponding to positive eigenvalues $\Lambda\approx
1$ were used to find \emph{almost-invariant sets}~\cite{DJ99,FD03,gary05,FPET07}.
The key point of difference between these prior studies and the present work is that
the sets studied previously do not move significantly over the time duration
studied, while
our present work seeks \emph{highly mobile} coherent sets, which are far from being almost-invariant. The new
theory and numerics we introduce in the next section are specifically designed for nonautonomous or
time-dependent systems.

\section{Numerical Approach}
We  partition $X$ into a grid of boxes $\{B_1,\ldots,B_n\}$. The
pressure extents of the boxes are either 3-5, 5-7, 7-10, or 10-20
hPa. Each pressure layer consists of 6605 boxes of approximately
equal cross-sectional area in the latitude/longitude directions,
leading to $n=26420$ boxes in total. To numerically approximate the
transfer operator $\mathcal{P}_{t,\tau}$, we construct a
finite-dimensional approximation based on Ulam's
approach~\cite{ulam}:
\begin{equation}\label{eq:measureP}
\mathbf{P}^{(\tau)}(t)_{i,j}=\frac{m(B_i\cap\Phi(B_j,t+\tau;-\tau))}{m(B_i)},
\end{equation}
where $m$ is a normalised volume measure in (lat,lon,pressure)
coordinates. The entry $\mathbf{P}^{(\tau)}(t)_{i,j}$ may be
interpreted as the probability that a point selected uniformly at
random in $B_i$ at time $t$ will be in $B_j$ at time $t+\tau$. The
discretisation naturally produces a diffusion at the level of box
diameters. As our boxes are of approximately the same dimensions as the distances between neighbouring ERA-40 data points, it is unnecessary to impose additional diffusion. If our boxes were significantly smaller the distances between neighbouring data points, it is possible that spurious fine features below the resolution supported by the data could appear;  in such a situation, additional diffusion would be required to remove spurious fine features.

We estimate $\mathbf{P}^{(\tau)}(t)_{i,j}$ by
\begin{equation}\label{eq:numericalulameqn}
\mathbf{P}^{(\tau)}(t)_{i,j}\approx{\#\{\ell:y_{i,\ell}\in B_i,
\Phi(y_{i,\ell},t;\tau)\in B_j\}}/{Q},
\end{equation}
where $y_{i,\ell}$, $\ell=1,\ldots,Q$ are uniformly distributed test
points in $B_i$ and $\Phi(y_{i,\ell},t;\tau)$ is obtained via a
numerical integration.
We set $Q=147$ in our experiments and calculate
$\Phi(y_{i,\ell},t;\tau)$ using the standard Runge-Kutta method with
step size of $3/4$ hours. Linear interpolation is used to evaluate
the velocity vector of a tracer lying between the data grid points
in the longitude-latitude-pressure coordinate. In the temporal
direction the data is affinely interpolated independently in the
longitude, latitude and pressure level directions. The step size of
$3/4$ hours is small enough to guarantee that a tracer will usually
not flow to a neighboring data grid set; this limits the numerical
integration error.

We assume that the mass density of particles in $X$ is at
equilibrium and denote the fractional mass of particles contained in
$B_i$ by $p_i$. Thus $\sum_{i=1}^n p_i=1$. We construct a reverse
time transition matrix from time $t+\tau$ to $t$ denoted
$\hat{\mathbf{P}}^{(\tau)}(t)$ as
$\hat{\mathbf{P}}^{(\tau)}(t)_{i,j}=\mathbf{P}^{(\tau)}(t)_{j,i}p_j/p_i$.

Introduce a weighted inner product $\langle x,y
\rangle_p:=\sum_{i=1}^n x_iy_ip_i$. One has $\langle
x\mathbf{P}^{(\tau)}(t),y \rangle_p=\langle
x,y\hat{\mathbf{P}}^{(\tau)}(t) \rangle_p$ for all $x,y\in
\mathbb{R}^n$.

Our new approach to finding a coherent pair is intuitively based
upon seeking a solution to \begin{equation}
\label{combprob}\max_{w\in\{\pm 1\}^n} \frac{\langle
w\mathbf{P}^{(\tau)}(t),w\mathbf{P}^{(\tau)}(t)\rangle_p}{\langle
w,w\rangle_p}.
\end{equation}

We think of $A_t:=\cup_{i:w_i=1}B_i$ and $A^c_t:=\cup_{i:w_i=-1}B_i$
as a coherent partition of $X$.   The numerator in (\ref{combprob})
represents the size of the forward image of the vector $w$.  If
there is little transport from $A_t$ to $\Phi(A^c_t,t;\tau)$ and
from $A^c_t$ to $\Phi(A_t,t;\tau)$ (so $A_t, \Phi(A_t,t;\tau)$ and
$A^c_t, \Phi(A^c_t,t;\tau)$ are both coherent pairs), this numerator
will be
large.  
To produce non-trivial partitions, we may need to place lower bounds
on the masses of both $A_t$ and $A_t^c$. Such a balanced bisection
problem is combinatorially hard to solve.  Therefore we remove the
discrete condition $w\in\{\pm 1\}^n$, allowing $w$ to float freely
in $\mathbb{R}^n$. To effect a balancing of mass between positive
and negative components of $w$, we insert the condition $\langle
w,v\rangle_p=0$, for some nonnegative test vector $v\in\mathbb{R}^n$.
We will see shortly that the correct choice of $v$ is the minimizer of the central inner product.
Thus, we have
\begin{equation}
\label{relaxprob}\min_{v\in \mathbb{R}^n}\max_{w\neq 0, \langle
w,v\rangle_p=0}\frac{\langle
w\mathbf{P}^{(\tau)}(t),w\mathbf{P}^{(\tau)}(t)\rangle_p}{\langle
w,w\rangle_p}.
\end{equation}
Letting $D_{ij}=\delta_{ij}p_i$ and noting $\langle
w,v\rangle_p=\langle wD^{1/2},vD^{1/2}\rangle_2$, this is easily
solved by computing the second largest singular value $s$ of
$D^{-1/2}\mathbf{P}^{(\tau)}(t)D^{1/2}$. Denote the corresponding
left singular vector by $y$ (under multiplication on the right). The
maximizing $w=w(t)$ is constructed as $w(t)=yD^{-1/2}$. The
minimizing $v$ turns out to be $uD^{-1/2}$ where $u$ is the leading
left singular vector of $D^{-1/2}\mathbf{P}^{(\tau)}(t)D^{1/2}$. We
also construct $z$ as the corresponding right singular vector and
set $w'(t+\tau)=zD^{-1/2}$. We assume that $w(t), w'(t+\tau)$ are
normalised so that $\langle w(t),w(t)\rangle_p=1$ and $\langle
w'(t+\tau),w'(t+\tau)\rangle_p=1$.

One now has:
\begin{enumerate}
\item $w(t)\mathbf{P}^{(\tau)}(t)=sw'(t+\tau)$,
\item $w'(t+\tau)\hat{\mathbf{P}}^{(\tau)}(t)=sw(t)$,
\item $\langle
w(t)\mathbf{P}^{(\tau)}(t),w(t)\mathbf{P}^{(\tau)}(t)\rangle_p=s^2$,
\end{enumerate}
Choosing $v$ via the minimization in (\ref{relaxprob}) ensures that $w$ has the transformation properties 1.\ and 2.\ above, which are crucial to the definition of coherent sets.

We now extract a coherent pair $A_t$ and $A_{t+\tau}$ from a pair of
vectors $w(t)$ and $w'(t+\tau)$. We create sets that are unions of
boxes with $w$-values above certain thresholds. Define
$\widehat{A}_t^+(c):=\bigcup_{i:w(t)>c}B_i$ and
$\widehat{A}_{t+\tau}^+(c):=\bigcup_{i:w'(t+\tau)>c}B_i$,
$c\in\mathbb{R}$. Denote
$\mu_n(\widehat{A}_t^+(c))=\sum_{i:w(t)>c}p_i$ and
$\mu_n(\widehat{A}_{t+\tau}^+(c))=\sum_{i:w'(t+\tau)>c}p_i$.

For $A_t=\bigcup_{i\in I_t} B_i$ and $A_{t+\tau}=\bigcup_{i\in
I_{t+\tau}} B_i$, define
$$\rho_n(A_t,A_{t+\tau})=\sum_{i\in I_t, j\in
I_{t+\tau}}p_i\mathbf{P}^{(\tau)}(t)_{ij}/\sum_{i\in I_t}p_i.$$ The
quantity $\rho_n$ measures the discretised coherence for the pair
$A_t, A_{t+\tau}$. Our procedure is summarised below:
\begin{enumerate} 

\item Let $\eta(c)=\arg\min_{c'\in\mathbb{R}}\bigl|
\mu_n(\widehat{A}_t^+(c))-\mu_n(\widehat{A}_{t+\tau}^+(c'))\bigr|$.
This is to enforce $\mu_n(A_t)=\mu_n(A_{t+\tau})$.
\item Set $c^*=\arg\max_{c\in\mathbb{R}}\rho_n(A_t^+(c),A^+_{t+\tau}(\eta(c)))$. The value of $c^*$
is selected to maximize the coherence. \item Define
$A_t:=\widehat{A}_t^+(c^*)$ and
$A_{t+\tau}:=\widehat{A}_{t+\tau}^+(\eta(c^*))$.
\end{enumerate}
We remark that one has to ensure that the sign of $w(t)$ and
$w'(t+\tau)$ manifest the same ``parity", i.e., the salient features
of $w(t)$ and $w'(t+\tau)$ to be extracted must have the same sign.
It may thus be necessary to multiply one of $w(t)$ or $w'(t+\tau)$ by $-1$.

The major computational cost is the construction of $\mathbf{P}^{(\tau)}$.
The calculation of large singular values and corresponding singular vectors is relatively quick, as $\mathbf{P}^{(\tau)}$ is very sparse and iterative methods for sparse matrices may be used.
The construction of $\mathbf{P}^{(\tau)}$ requires numerical integration of $Q\cdot n$ trajectories for a flow duration of $\tau$ time units.
The trajectory computations are of course highly parallisable, and further computational savings might be made by reusing already computed trajectory segments to link with new trajectories when the latter pass nearby.

\section{Numerical Results}
\begin{figure}[htb]
\centering
    \includegraphics[width=0.5\textwidth]{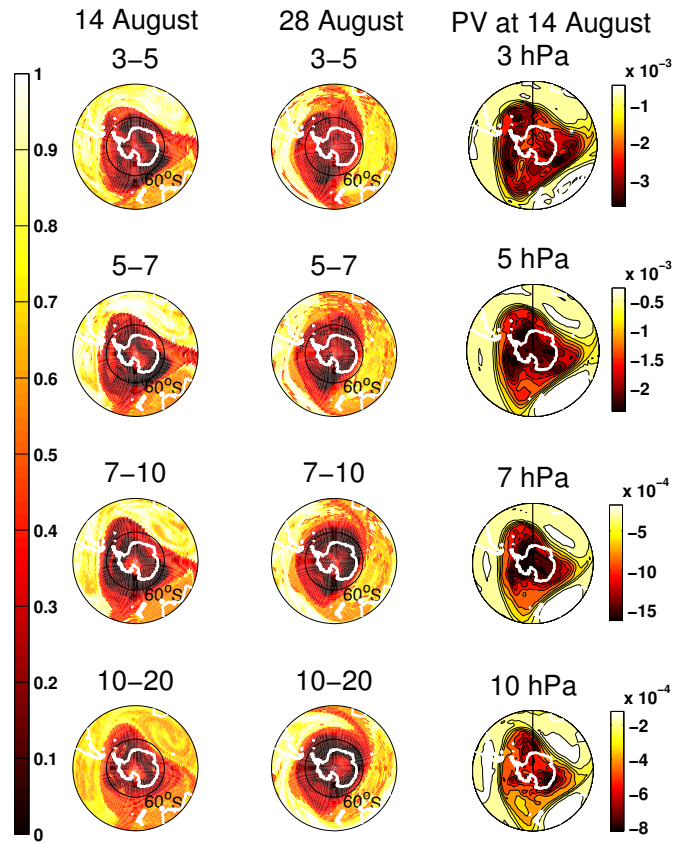}
    \caption{\label{fig:svPV}\footnotesize{Left column: the vector $w(14)$ shown for the pressure levels $3-5$, $5-7$, $7-10$ and $10-20$
    hPa. The $4\times 6605=26420$ components of $w(14)$ have been mapped
    to the values 1/26420,2/26420,\ldots,1, preserving
    their order.
    Center column: $w(28)$.
    Right column: Potential vorticity ($Km^2kg^{-1}sec^{-1}$) at levels
    3, 5, 7 and 10 hPa on August 14, 1999 ($t=14$).}}
   
\end{figure}

\begin{figure}[htb]
\centering
    \includegraphics[width=0.5\textwidth]{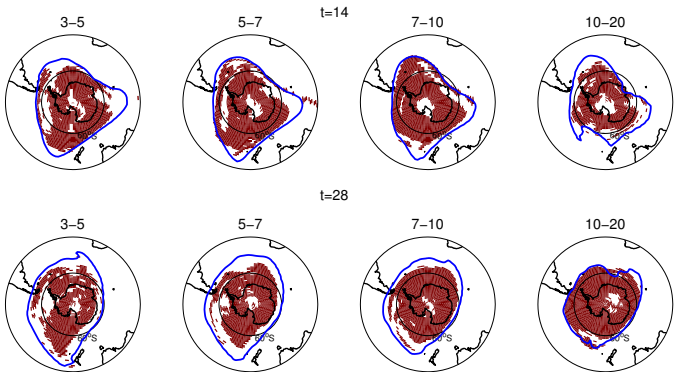}
     \caption{\label{fig:coherentsetandPV}\footnotesize{Comparison between coherent pair and PV surface boundary.
    The optimal coherent set $A_{14}$ and $A_{28}$ obtained from thresholding the vectors $w(14)$ and $w'(28)$. The coherent ratio $\rho_{\mu}(A_{14},A_{28})\approx$ 0.7902. The blue curve in each plot shows the PV surface boundary obtained from the maximum PV gradient w.r.t equivalent latitude}}
\end{figure}

We computed the SVD of $D^{-1/2}\mathbf{P}^{(t)}(\tau)D^{1/2}$ at
$t=14$ with $\tau=14$ days to obtain the left (resp.\ right)
singular vectors $y$ (resp.\ $z$) and hence $w(14)$ and $w'(28)$.
Figure~\ref{fig:svPV} illustrates the vectors $w(14)$ and $w'(28)$ with the components monotonically rescaled to uniformly distributed values between 0 and 1.
The highlighted part of these vectors describes the most coherent
pair of sets.  We now threshold $w(14)$ and $w'(28)$ using the
algorithm described above to extract the corresponding pair of
coherent sets; see Figure~\ref{fig:coherentsetandPV}. We find the
optimal coherence ratio is $\rho_n(A_{14},A_{28})\approx 0.7902$;
this means that about 21\% of the mass in $A_{14}$ on August 14,
1999 falls outside $A_{28}$ on August 28, 1999.

Interestingly, our
coherent pair has a ``hole'' over the south pole.  Further
calculations have revealed that the reason for this is that around
twice as many particles in this vertical hole on August 14 have
exited the slice 3-20 hPa vertically by August 28 when compared to
similar vertical exits of particles starting in the identified
coherent set on August 14. Thus, this inner part of the vortex is
less coherent and excluded from our coherent pair. This hole may be an artifact of the reanalysis data, although it is consistent with evidence of very strong polar descent in this region \cite{Manney94}.


We now compare our coherent pair of sets to sets defined by contours
of potential vorticity (PV).
A common approach, developed
in~\citep{McIntyre83, NNRS96} is to define the vortex boundary as
the isoline of the largest gradient of PV w.r.t. the
\emph{equivalent latitude}. We employ this approach to define
potential coherent pairs at $t=14$ and $t=28$.  We additionally
enforce the constraint that the mass enclosed by a PV isocontour at
$t=28$ is approximately equal to the mass of the set enclosed by the
determined PV isocontour at $t=14$.
The computational cost of the PV approach is NARATIP, PLEASE ADD MATERIAL ON COMPUTATIONAL COST.

The two-dimensional plots of PV-determined coherent pairs at $t=14$
and $t=28$ are compared with the coherent sets in
Figure~\ref{fig:coherentsetandPV}. To estimate the transport of
particles from the inside the set at $t=14$ to outside the set at
$t=28$, we use a method similar to the contour crossing method
introduced in~\cite{Sobel97}. The tracer particle is considered to
be outside the boundary if its potential vorticity is larger than
that of the boundary. Note that the contour crossing method is
originally developed to estimate the transport on the 2D isentropic
surface but we would like to extend its utility to estimate the
transport across the boundary surface. Therefore, we interpolate the
PV at the final time ($t=28$) to obtain the PV at the particle's
final position. We also interpolate the PV of the boundaries of the
set at $t=28$ along the pressure coordinate to determine the
boundary at the pressure level the advected particle resides in.
This calculation shows that the fraction of particles initially
inside the surface at $t=14$ remains inside the boundary surface at
$t=28$ is approximately 0.7204. Thus our transfer operator based
approach yields coherent pairs with 9.69\% greater coherence.
Moreover, our approach is able to detect finer structures, including
a vertical hole near the south pole.

\section{Conclusions}

The Antarctic polar vortex is a well-known feature of the austral
wintertime stratosphere separating polar and midlatitude air masses.
The strong barrier to transport at the vortex edge plays an
important role in ozone dynamics, particularly the development of
the Southern Hemisphere ozone hole in austral spring. Diagnosis of
the vortex edge from observations is a challenging problem that
remains a subject of active research.

Previous approaches to this problem have been based on kinematic
(following the advection of some tracer) or dynamic (considering
gradients of PV) arguments. We presented a new kinematic method of
accurately estimating the three-dimensional location of the vortex.
This new
  method uses the velocity field to diagnose ``optimally coherent pairs'' and was able to determine a significantly more accurate estimate
of transport barriers, with almost 10\% less external transport from
the identified vortex region than the PV-based estimate.
Future, more detailed studies will include an investigation of the climatology of the polar vortex on isentropic surfaces throughout the stratosphere.

Our new computational approach for detecting
minimal transport structures with high accuracy has a broad range of
potential application to studies of transport and mixing in the
atmosphere and ocean, and in general fluid dynamics settings.

\begin{acknowledgments}
GF is partially supported by the ARC Centre of Excellence for the
Mathematics and Statistics of Complex Systems (MASCOS).  NS is
supported by a MASCOS fellowship. AM is supported by NSERC.
\end{acknowledgments}

\end{document}